\documentclass[aps,prl,twocolumn,showpacs,groupedaddress,amssymb]{revtex4}
\usepackage{graphicx}
\usepackage{color}
\usepackage{bm}
\usepackage{doi}
\usepackage{float}
\usepackage{amsmath}

\newcommand{\rb}[1]{\color{blue} #1}

\begin{document}
	
	\title{Exceptional Point Generated Robust Asymmetric High-Order Harmonics}
	\author{Gui-Lei Zhu$^1$}
	\thanks{These two authors contributed equally}
	\author{Amir Targholizadeh$^1$}
	\thanks{These two authors contributed equally}
	\author{ Xin-You L\"{u}$^2$}
		\email{xinyoulu@hust.edu.cn}
	\author{ Cem Yuce$^3$}
	\author{Hamidreza Ramezani$^1$}
	\email{hamidreza.ramezani@utrgv.edu}
	
	\affiliation{$^1$ Department of Physics and Astronomy, University of Texas Rio Grande Valley, Edinburg, Texas 78539, USA\\
		$^2$ School of Physics, Huazhong University of Science and Technology, Wuhan 430074, China\\
		$^3$ Department of Physics, Eskisehir Technical University, Eskisehir, Turkey}

	\begin{abstract}
		We propose a metallic-silicon system with a complex optical potential modulated along the length of the waveguide for a robust higher harmonic generation. For right moving fields when the strength of non-Hermiticity becomes equal to the real part of the optical potential, the dynamical equations associated with modal field amplitudes in our proposed system are described by a Jordan form Hamiltonian. This ultimately will allow for a unidirectional higher frequency generation which always has a maximum value for a specific length of the waveguide irrespective of the geometrical imperfections in the design of the waveguide. Furthermore, the phase of the generated higher harmonic mode becomes independent of the coupling between the fundamental frequency and higher harmonic one. Unlike other proposed spatiotemporal modulated systems when the system has a Jordan form Hamiltonian, the fundamental mode remains reciprocal while the harmonic generation is non-reciprocal. Consequently, while the proposed device cannot be used as an optical isolator it can be used for many other devices such as laser cavities, interferometry, and holographic processes. 
		
	\end{abstract}

	\maketitle
	{\it Introduction---} Optical metamaterials with complex potentials provide incredible possibilities to alter light-matter interaction, create new functionalities and artificially change optical properties of synthetic structures that are not possible with normal materials\cite{1,2,3,4,5}. Among such fascinating effects in light-matter interactions are unidirectional invisibility\cite{6,7}, topological lasers\cite{8,9}, isolation and circulations\cite{10,11,19,fn,alus, NS}, new lasing schemes\cite{12,13}, low-cost quantum state transfer\cite{14,15,fm20} and quantum lattice with multi-point phase transition\cite{GL20}, one and two dimensional wave manipulation\cite{16,17,XX}, to name a few. These new functionalities are achieved using different physical mechanisms. For example, in a parity-time symmetric system, the major element is non-Hermiticity where some gain and loss mechanisms break the Hermiticity in a system and results in the bi-orthogonality of states. A consequence of non-Hermiticity is the existence of exceptional points in which the Hilbert space becomes skewed and two or more eigenstates coalesce\cite{18,19,20,21,22,23,24}. In general, the exceptional points are not robust and are affected by the other parameters in the system. However, there are reports for the generation of a robust exceptional point where the underlying physics is associated with the nilpotent matrices with an odd number of sites\cite{25}. 
	
	The other example of metamaterials is the time modulated system and specifically the combination of space and time modulation\cite{26,27,28,29,30,rmprl17}. The space and time modulated systems are known for breaking reciprocity, and generation of optical isolators based on asymmetric harmonic generation. Time modulated systems have been studied in waveguides, ring resonators, and metasurfaces\cite{31,32,33}. In more recent studies, the spatiotemporal modulation has been extended to complex potentials with gain and loss mechanisms where higher-order knots and asymmetric exceptional points have been reported\cite{34}. However, the interplay between the exceptional point and harmonic generation has not studied yet. Nonlinear harmonic generation (NHG) is a beneficial procedure for coherently producing laser light in the extreme-ultraviolet region of the spectrum\cite{35,36,37}. The applications include: investigating surface dynamics\cite{38,39}, holographic imaging\cite{40,41}, probing static molecular structure, or internal molecular dynamics\cite{42,43}.

	One of the main challenges in harmonic generation in laser cavities is the susceptibility of the output emission to the fabrication process and low-efficiency of the NHG. Although recently there have been some attempts to overcome the efficiency issue, there are problems in terms of sensitivity of the structure to the fabrication process, losses, and frequency restrictions\cite{44,45,46}.

	This paper investigates an alternative method to address the aforementioned difficulty regarding fabrication process and achieve robust higher harmonic generation using time and space modulation of the real and imaginary part of the refractive index. Specifically, we show that an appropriate selection of a spatially and temporally varying modulation of the refractive index causes simultaneous changes in the frequency and momentum of photons traveling in the waveguide allowing a robust NHG. The specific key element in this modulation is the concurrent spatiotemporal modulation of the real and imaginary part of the distributed relative permittivity with equal strength. This specific modulation makes the system operate at the exceptional point associated with the coupled equations governing the dynamics of the modes.

	Exceptional points are particular singularities that arise in non-Hermitian complex systems, in which two or more eigenvalues and eigenstates coalesce. As a result of the coalescence of the eigenvectors at the exceptional point, the system undergoes a phase transition from a real spectrum to a complex one. Initially, exceptional points have been proposed in spatially modulated systems and demonstrated a rich physics and depict fascinating features specifically in the so-called parity-time symmetric systems with balanced amplification and absorption mechanisms.  Among these intriguing features and applications are unidirectional physics and enhanced sensitivity, which found their ground in optics\cite{6,47,48,49,50}.

	Here our modulation in the real and imaginary part of the index of refraction is not restricted to space and encompasses the time-domain too. Such non-Hermitian modulation can result in the generation of exceptional points associated with the quasi energies that have a dependency on the direction of excitation\cite{11,34}. In contrast to this interesting feature, here we are focusing on the dynamical properties of the system with a non-Hermitian spatiotemporal modulation and show that a robust NHG can be obtained. This could be a new direction of using exceptional points to engineer devices that are robust against undesirable imperfections. 
	
		\begin{figure}
		\includegraphics[width=0.98\linewidth, angle=0]{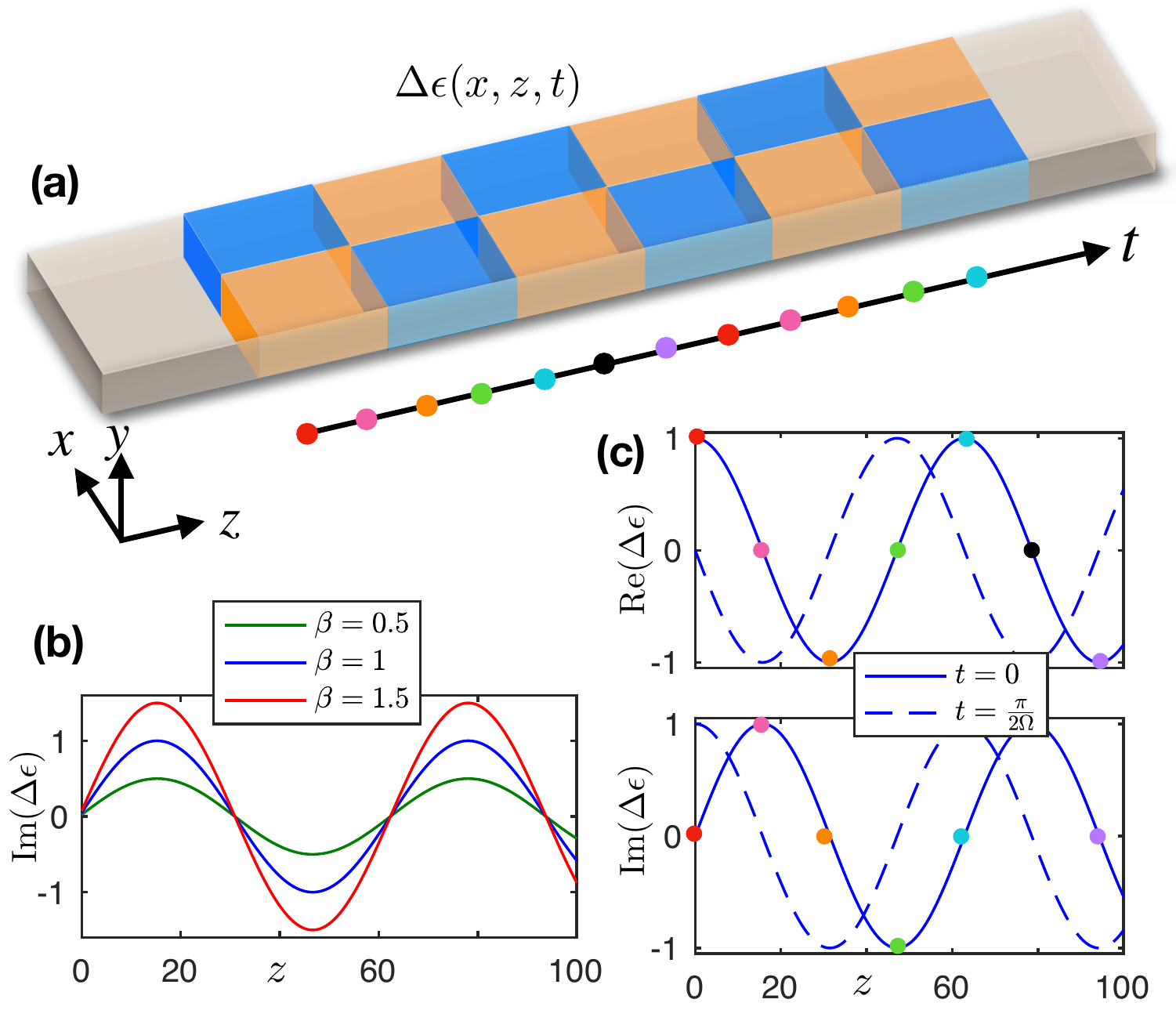}
		\caption{(Color online) (a) Schematic diagram of the proposed system. The wavaguide is modulated by periodical space and time-dependent complex optical potentials [see Eq.\,(\ref{1})]. The blue and orange parts represent the real and imaginary of $\Delta \epsilon$, respectively. (b) The imaginary part of $\Delta \epsilon$ with different $\beta$. Factor $\beta$ changes the amplitude of imaginary part of the $\Delta \epsilon$ where we consider $\delta(x)=1, q=0.1, \Omega t=0$. (c) The real and imaginary parts of $\Delta \epsilon$ versus different time for $\beta=1$. The dots with different color represent varying time, which is related to the $t$ axis in (a).}
		\label{fig1}
	\end{figure}
	
	{\it Theoretical analysis---} To demonstrate robust second-harmonic generation, let us consider an isolated silicon waveguide. The dispersion of the waveguide and its associated modes are governed by the boundary conditions and define the permitted electromagnetic waves to propagate inside the waveguide.  In a linear system with no time modulation, the modes are perpendicular to each other, and no transition occurs between the two modes. Whereas, in a spatiotemporal modulated system under a certain condition, a transition can occur between the two modes resulting, for instance, to NHG. The strength of the mode conversion is sensitive to the coupling between the two modes. The coupling between the two modes is directly defined by the geometrical parameters of the waveguide. Consequently, any changes in the geometrical properties of the waveguide, specifically the width of it, can change the NHG. Our major aim here is to address this challenge and propose a structure that can produce a robust NHG which is immune to geometrical flaws.
	
	The waveguide that we consider here has a space-time-dependent permittivity distribution along the propagation direction of the waveguide which is described by $\epsilon_w (x,z,t)=\epsilon_s+\Delta \epsilon(x,z,t)$. Here $\epsilon_s$ is silicon relative permittivity constant that in our numerical calculation we consider being $12.25$. Furthermore, $\Delta\epsilon$ is our proposed distributed relative permittivity that is given by:
	\begin{equation}
		\Delta\epsilon(x,z,t)=\delta(x)\left[\cos (\Omega t+qz)+i\beta \sin (\Omega t+qz)\right],
		\label{1}
	\end{equation} 
	where $\delta(x)=\delta$ is a zero constant outside of the waveguide and a nonzero one inside it, describing the modulation amplitude along the $x$-direction. In this medium, the imaginary part of the modulation denotes the gain and loss depending on the sign of it in Eq.(\ref{1}). By altering the system parameter $\beta$, we can control the strength of imaginary part modulation or in other words the gain and loss parameter in our system, which ultimately dictates the PT-phase transition. An example of such permitivity is shown in Fig.\,\ref{fig1}.

\begin{figure}
		\includegraphics[width=1\linewidth, angle=0]{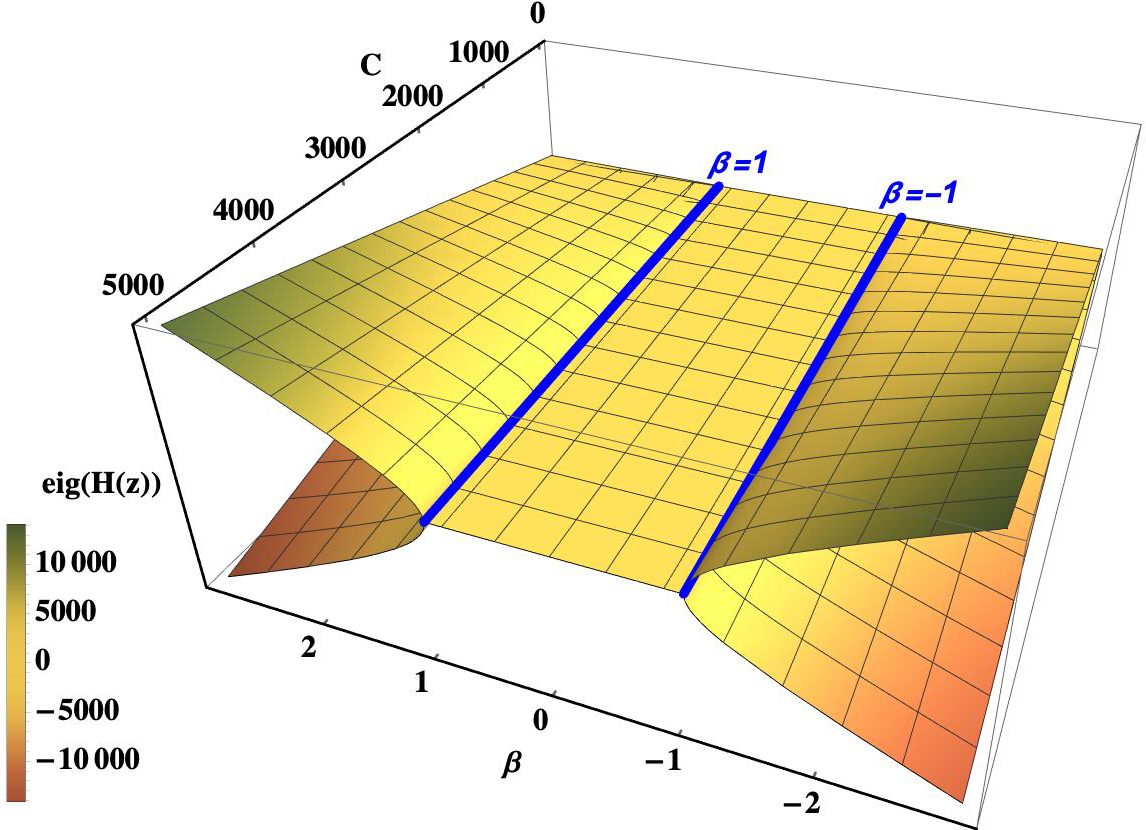}
		\caption{The real part of the Hamiltonian's eigenvalue in Eq.(\ref{3}) is depicted as a function of $ C $ and $ \beta $. We can observe that when $ \beta=1, -1 $ eigenvalues coalesce; thus, we have exceptional point. Interestingly, the value of $ C $ doesn't change the location of exceptional points if $ \beta=1, -1 $. In other words, exceptional poiint is robust against geometrical imperfections. }
		\label{fig2}
	\end{figure}

	The non-perturbed silicon waveguide possesses many bands; here, we only consider the first two modes of the waveguide that are coupled to each other with the highest efficiency in the presence of the time-dependent modulation. The first mode is a symmetric mode at frequency $\omega_1$ with wave vector $k_1$ and the second one is the anti-symmetric mode with frequency $\omega_2$ with the wave vector $k_2$. To continue with the analytical calculations, we consider the modulation frequency to be given by $\Omega=\omega_2-\omega_1$ with $\Delta k=q+k_1-k_2$. The phase-matching condition, in this case, is given by $\Delta k=0$, however, we do not restrict ourselves to operating under the phase-matching condition. Note that our proposal is reduced into a time-independent one when $\Omega=0$, which has received much attention, especially in realizing unidirectional invisibility \cite{6}. By considering only two propagating modes toward the right, the electric field in the waveguide can be written as
	\begin{equation}
		E(x,z,t)=\sum_{n=1}^{2}a_n(z)E_n(x)e^{i(\omega_n t - k_n z)}.
		\label{2}
	\end{equation}
	Above $E_{1,2} (x)$ are the modal profiles and $\left|a_{1,2}\right|^2$ are the corresponding normalized intensities associated with each mode. Substituting $E(x,z,t)$ and Eq. (\ref{1}) into Maxwell equations and applying the slowly varying envelope approximation we obtain the coupled-mode equations for the modal field amplitudes $\vec{\Psi}(z)=\left(a_1(z)\quad a_2(z)\right)^T$
	\begin{equation}
		i\frac{d}{dz}\vec{\Psi}=H(z)\vec{\Psi}
		\label{3}
	\end{equation}
	in which
	\begin{equation}
		H(z)=\left(\begin{array}{cc}
			0&C(1-\beta)e^{-i\Delta kz}\\
			C(1+\beta)e^{i\Delta kz}&0
		\end{array}\right),
	\end{equation}
	where $C=\frac{\epsilon_0}{8}\int_{-\infty}^{\infty}E_1(x)\delta(x) E_2(x)dx$ is the coupling strength. It is clear that the integral has nonzero values inside the waveguide and the width of the waveguide directly changes the interval of the integral; thus affects the coupling $C$. However, we observe that when $\beta=\beta_r=1$ the matrix $H(z)$ becomes a Jordan form and a nilpotent matrix of the second order since $H ^2 = 0$. Therefore, as it is shown in Fig.\,\ref{fig2}, $\beta_r$ can be considered as the second order exceptional point of the $H(z)$, irrespective of the value of $C$ or $z$. Specifically, for $\beta_r=1$, the intensity of symmetric mode, namely $a_1$ remains constant while $a_2$ would change through the evolution along the $z$ direction. 

	\begin{figure}
		\includegraphics[width=1\linewidth, angle=0]{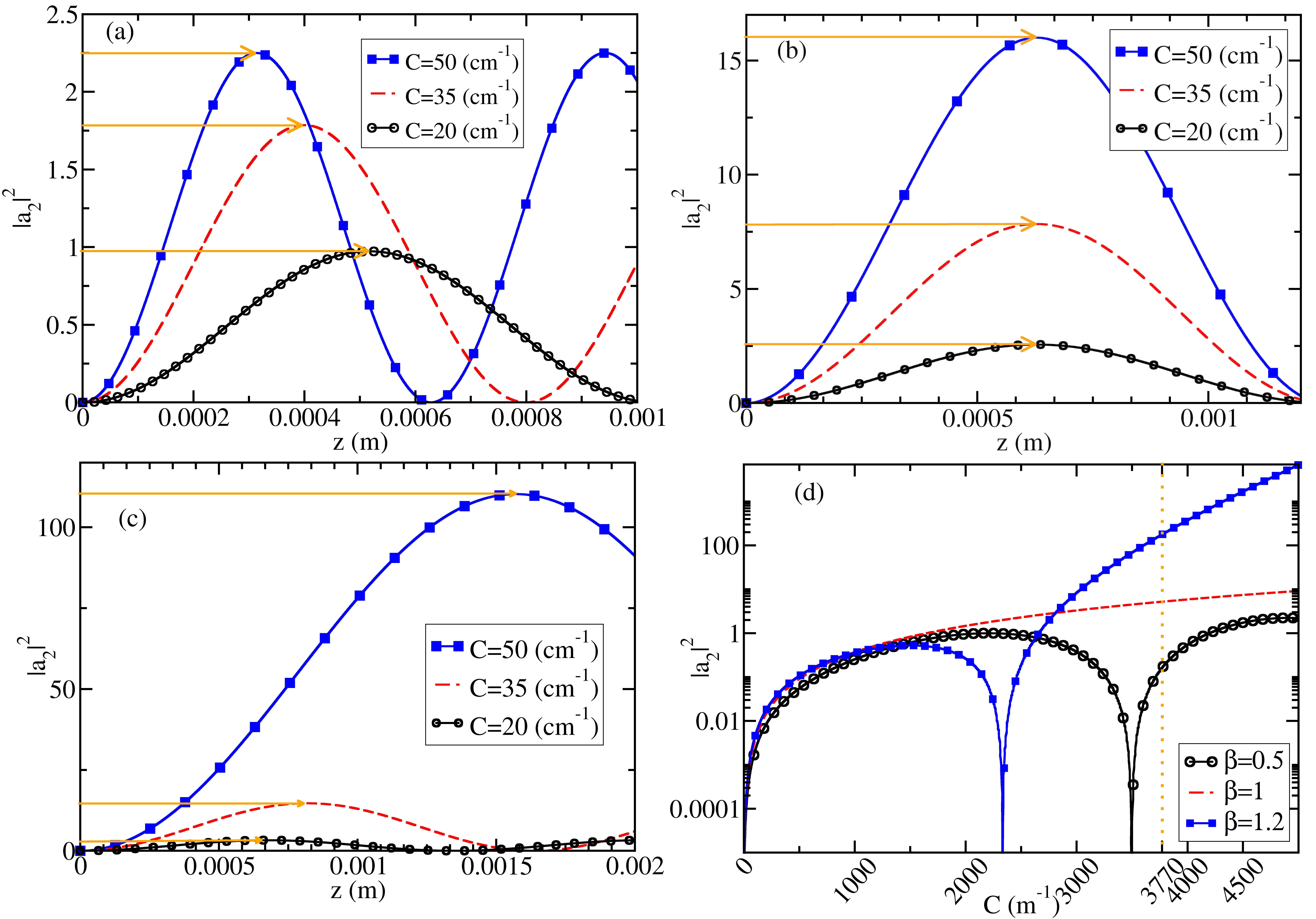}
		
		\caption{(a),(b),(c) Intensities of NHG are depicted for  $\beta=0.5 (a), 1 (b), 1.1 (c)$ as a {\rb f}unction of waveguide's length for three different value of $ C $. Arrows are showing the length which  NHG's {\rb i}ntensity is maximum on that. Clearly, this length, is associated with the value of $ C $ when  $\beta=0.5, 1.1$. while for $\beta=1$ it is independent from $ C $.(d)  Log-linear plot of intensities of NHG are depicted for  $\beta=0.5, 1, 1.2$ as a Function of $ C $ at the end of a waveguide with specific length. We also have identified the value of {\rb $C$} for which the exceptional point occurs with a vertical dash line for $\beta=1.2$. Even in the exact phase for $\beta\neq 1$ the changes in $C$ result in significant changes in $a_2$. However at $\beta=1$ it is linearly proportional to $C^2$.}
		\label{fig3}
	\end{figure}

	We can analytically solve coupled differential equations in Eq.(\ref{3}) for any values of $\beta$. For example, for the initial excitation of symmetric mode from the left side of the waveguide namely $\vec{\Psi}(0)=\left(a_1 (0)\quad a_2 (0)\right)^T=\left(1\quad 0\right)^T$ we would have 
	\begin{equation}
		\left(\begin{array}{c}
			a_1(z)\\
			a_2(z)
		\end{array}\right)=\left(
		\begin{array}{c}
			e^{-\frac{iz\Delta k}{2}} \left[\cosh\frac{ z\zeta}{2}+\frac{i\Delta k \sinh\frac{z\zeta}{2}}{\zeta}\right]\\
			-2iCe^{\frac{iz\Delta k}{2}} (1+\beta)\frac{\sinh\frac{z\zeta}{2}}{\zeta}
		\end{array}\right),
		\label{5}
	\end{equation}
	in which $\zeta= \sqrt{4C^2(\beta^2-1)-\Delta k^2}$. From Eq.(\ref{5}) it is clear that for $\beta=\beta_r$, $\vec{\Psi} (z)|_{\beta=\beta_r}$ will be given by 
	\begin{equation}
		\vec{\Psi} (z)|_{\beta=\beta_r}=\left(\begin{array}{c}
			a_1(z)\\
			a_2(z)
		\end{array}\right)=\left(
		\begin{array}{c}
			1\\
			-\frac{4iC}{\Delta k}  \sin(\frac{\Delta k z}{2})e^\frac{iz\Delta k}{2} 
		\end{array}\right).
		\label{6}
	\end{equation}
	Before we make further discussion on the generated harmonic amplitude $a_2$ we would like to bring to the attention of the reader that $a_1$ remains constant and is not affected by the time dependent parity-time symmetric modulation. This effect reminds us about unidirectional invisibility effect when $\Omega=0$\cite{6}. 
	
	From relation (\ref{6}) it is obvious that for $\beta_r$ the maximum of the amplitude of the generated higher harmonic term $\left|a_2 (z)\right|^2$ always occurs at length $z=\frac{(2m+1)\pi}{\Delta k}$ where $m$ is an integer number. However, if $\beta\neq\beta_r=1$ then the maximum of the intensity of generated higher harmonic $\left|a_2 \right|^2$ depends on the value of $C$. In practical applications, the ideal length for designing a waveguide would be a length that the intensity of the generated harmonic finds its maximum. However, in other harmonic generation methods this length is affected by the coupling $C$. Thus, it is an extremely advantageous aspect of our proposal where this length becomes independent of the value of $C$. This conclusion is depicted in Fig. \ref{fig3} where we plotted the $\left|a_2 \right|^2$ as a function of $z$ for $\beta=0.5, 1, 1.1$ in panel (a-c), respectively. In each panel we considered $\Delta k=50$ $\rm{cm}^{-1}$ and $C = 20 ,35, 50$ $ \rm{cm}^{-1}$. Notice that for $\beta>\sqrt{1+(\frac{\Delta k}{2C})^2}$ the system undergoes a phase transition to the broken phase, and it becomes unstable. The $\beta=\beta_{{\rm{EP}}}=\sqrt{1+(\frac{\Delta k}{2C})^2}$ is the exceptional point defined by the quasi-energies $\pm{\cal E}$ in the system\cite{34}. This can be seen by applying Floquet theory on Eq. (\ref{2}), where we find the quasi-energy of the system as
	\begin{equation}
		{\cal E}_\pm=\frac{\Delta k}{2}\pm\sqrt{(\frac{\Delta k}{2})^2-C^2 (\beta^2-1))}.
	\end{equation}
We can easily show that at the exceptional point $\beta_{{\rm {EP}}}$ associated with quasi energies the $\left|a_2\right|^2$ increases in power low manner namely it becomes proportional to $z^2$. Thus, $\beta=\beta_r$ is always a stable point for the system to operate in it.

	In  Fig. \ref{fig3}(d), we plotted the intensity of $a_2$ for a specific length $z=1.6$ mm as a function of $C$ for three values of $\beta$. We have identified the value of $C$ for which the exceptional point occurs with a vertical dash line for $\beta=1.2$. Note that the exceptional point as a function of $C$ is given by $\sqrt{\frac{\Delta k ^2}{4(\beta^2-1)}}$and thus only for $\beta>1$ we can find a $C$ that makes the system to operate at the exceptional point. It is clear that even when we are in the exact phase for $\beta\neq\beta_r$ the changes in $C$ result in significant changes in $a_2$ while at $\beta=\beta_r$ it is linearly proportional to $C^2$.

\begin{figure}
		\includegraphics[width=1\linewidth, angle=0]{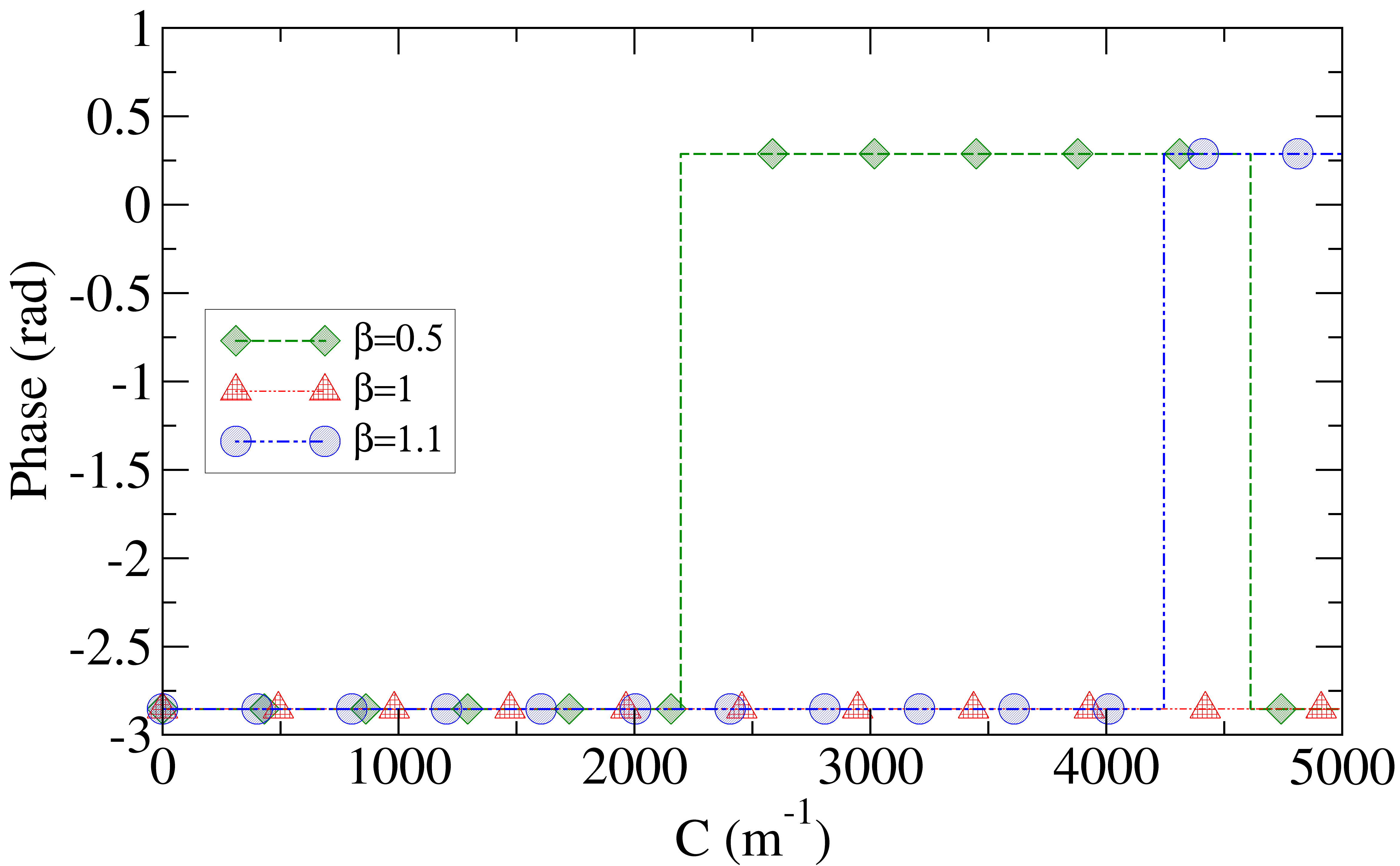}
		\caption{Phase of $ a_2 $ is represented as a function of $C$ for three different values of $ \beta $ at the end of a waveguide with specific length $z=2$ mm. This phase changes the electric field phase in Eq.(\ref{2}). clearly, for $\beta\neq 1$ output field will experience sudden changes in its phase respected to $C$. In contrast when $\beta=1$ only the trivial phase due to the phase-mismatching condition will be added to the output field.    }
		\label{fig4}
	\end{figure}

	In a dynamical system, intensity is not the only important element. Another important quantity is the phase. Specifically, in different applications such as designing a laser cavity based on an NHG, in the interferometry processes, and in Holographic imaging the phase of the output field is crucial in the device performance. According to Eq.\,(\ref{5}) any mechanism that affects the phase of $a_2$ results in changing the phase of NHG in output. From Eq. (\ref{6}) it is clear that the phase of output amplitude $a_2$ is independent of $C$ when $\beta=\beta_r$. However, for other values of $\beta$ as long as we are in the exact phase, the phase of the $a_2$ changes by $\pm\pi$ depending on the value of $C$. To demonstrate this, in figure \ref{fig4} we plotted the changes in the phase of the output amplitude $a_2$ for three values of $\beta$ for a specific length of a waveguide, namely $z=2$ mm as a function of $C$. 
	
	
	 All the above discussions are valid when we excite the system from the left side and the direction of propagation is toward the right. On the other hand, we know that usually, the spatiotemporal modulations result in optical isolation. For the right incident excitation and at $\beta=\beta_r=1$ it is easy to show that the Hamiltonian associated with Eq.(\ref{3}) reduces to a $2\times 2$ zero matrix and thus $\left(a_1(z)\quad a_2(z)\right)=\left(a_1(0)\quad a_2(0)\right)$ and thus not only is not any harmonic  generated by right excitation but also the system is reciprocal even we have asymmetric transport. This means that at $\beta=\beta_r$ we do not have any optical isolation.

	{\it Conclusion---} We have proposed a new application for exceptional points in complex spatiotemporal modulated systems where they allow for a robust and efficient harmonic generation. While the non-reciprocity becomes suppressed at the exceptional point of the dynamical system, the robust harmonic generation becomes valuable specifically in applications such as interferometry, Harmonic generation-based lasers, and holography applications. 
	
\begin{acknowledgments}
		H. R. acknowledge the support by the Army Research Office Grant No. W911NF-20-1-0276 and NSF Grant No. PHY-2012172. The views and conclusions contained in this document are those of the authors and should not be interpreted as representing the official policies, either expressed or implied, of the Army Research Office or the U.S. Government. The U.S. Government is authorized to reproduce and distribute reprints for Government purposes notwithstanding any copyright notation herein. 
		\end{acknowledgments}

\end{document}